\documentclass[twocolumn]{aa}
\usepackage{graphicx}
\usepackage{txfonts}

\def\ltsima{$\; \buildrel < \over \sim \;$}
\def\simlt{\lower.5ex\hbox{\ltsima}}
\def\gtsima{$\; \buildrel > \over \sim \;$}
\def\simgt{\lower.5ex\hbox{\gtsima}}
\begin{document}
   \title{The age of the main population of the Sagittarius dwarf spheroidal
          galaxy}

   \subtitle{Solving the ``M giant conundrum''.}

   \author{M. Bellazzini\inst{1}, M. Correnti\inst{2}, F.R. Ferraro\inst{2},
           L. Monaco\inst{3}
          \and
          P. Montegriffo\inst{1}\fnmsep\thanks{Based on observations collected at the European Southern 
	  Observatory, Chile, (Programme 71.D-0222A).}
          }

   \offprints{M. Bellazzini}

   \institute{INAF - Osservatorio Astronomico di Bologna,
              Via Ranzani 1, 20127, Bologna, ITALY \\
              \email{michele.bellazzini, paolo.montegriffo@bo.astro.it}
         \and
             Dipartimento di Astronomia, Universit\`a di Bologna,
	     Via Ranzani 1, 20127, Bologna, ITALY \\
             \email{francesco.ferraro3@unibo.it, matteo.correnti@studio.unibo.it}
         \and
	     ESO - European Southern Observatory, Alonso de Cordova 3107, 
	     Santiago 19, CHILE;
	     \email{lmonaco@eso.org}
             }

   \date{Accepted by Astronomy \& Astrophysics Letters}

   \abstract{We present a statistically decontaminated Color 
   Magnitude Diagram of a $1\degr\times 1\degr$ field in the core 
   of the Sagittarius dSph galaxy. Coupling this CMD with
   the most recent metallicity distributions obtained from high resolution
   spectroscopy we derive robust constraints on the mean age of the stellar
   population that dominates the galaxy (Pop A). Using
   three different sets of theoretical isochrones in the metallicity range
   $-0.4\le [M/H]\le -0.7$ and taking into consideration distance moduli
   in the range $16.90\le (m-M)_0\le 17.20$ we find that the mean age of Pop A
   is larger than 5 Gyr, and the best-fit value is age$ = 8.0\pm
   1.5$ Gyr. Since Pop A provides the vast majority of the M giants that traces
   the tidal stream of Sgr dSph all over the sky, our estimate resolves the so
   called ``M giant conundrum'' (Majewski et al. 2003). The time needed by
   the M giants that currently populates the stream to diffuse within the main 
   body of Sgr and to reach the extremes of the tidal tails once torn apart 
   from the parent galaxy ($\simeq 3-4$ Gyr) can be easily accommodated into 
   the time lapsed since their birth ($\simeq 5.5-9.5$ Gyr).

   }
    
   \authorrunning{M. Bellazzini et al.}
   \titlerunning{The age of the Sagittarius dSph} 
    
   \maketitle
%

\section{Introduction}

The Sagittarius dwarf spheroidal galaxy (Sgr dSph; Ibata et al. \cite{iba1})
currently provides the cleanest possibility to study in detail the tidal
disruption and the accretion of a dwarf satellite into a large galaxy. The
tidal tails of the disrupting galaxy have been observed in widely different
positions with different tracers (see, for instance, Newberg et al.
\cite{sdss}, Ibata et al. \cite{iba2}, and references therein).
In particular, Majewski et al. (\cite{maj03}, hereafter M03) showed that M
giants traces the tidal tails of Sgr as a coherent and  dynamically cold
filamentary structure (hereafter Sgr Stream) extending for tens of kpc from the
parent galaxy and nicely aligned along the rosetta orbit of  Sgr (see also Law,
Johnston \& Majewski \cite{law}). 

The study of the physical properties (age and chemical abundances)  and the
kinematics of stars into the main body of Sgr and into the Stream may provide
for the first time the possibility to link the Star Formation History (SFH) 
of the
galaxy with its ``dynamical/orbital'' history and evolution (see Bellazzini,
Ferraro \& Buonanno \cite{BFBb}, hereafter BFBb) and/or to constrain the  one
with the other. A first relevant case of this interlacing was pointed
out by M03. These authors obtained rough estimates of the metallicity of M
giants in the Sgr Stream from their J-K colors and using the Age-Metallicity
Relation (AMR) for Sgr provided by Layden \& Sarajedini (\cite{ls00}, hereafter
LS00), they concluded that Stream M giants should be younger than 5 Gyr and a
significant fraction of them have an age of 2-3 Gyr. The N-body models that
best reproduces the Sgr galaxy + Stream system (Law et al. \cite{law}) show
that a comparable amount of time is needed to produce tidal tails of the
observed extension. According to these models, we are observing in the Stream M
giants that were torn apart from their parent galaxy  up to 3.2 Gyr ago.
Since it is reasonable to imagine that in a spheroidal system star formation
episodes occur preferentially toward the center of the galaxy, the above time
lapse should be (possibly) increased by the time needed to the newly born 
stars to diffuse out to the ``edges'' of the system (M03).  
Hence, the comparison  between the stellar evolution
timescales (e.g., the age of M giants) and the dynamical timescales (e.g., the
time needed to populate the whole extension of the Stream) give rise to 
a possible inconsistency, or at least a fine-tuning problem, since stars 
cannot be torn apart from a galaxy before their birth.
M03 dubbed this apparent mismatch between the two timescales as "the M
giants conundrum''.

 However, the method adopted by M03 to estimate the age of Sgr M giants
is prone to uncertainties both in the previously established Sgr AMR
as well as in the predicted colors of RGB tip stars from theoretical
isochrones.
An age estimate 
based on more reliable age-sensitive observables (such as the Main Sequence 
Turn Off - TO -  and/or the Sub Giant Branch, SGB) would greatly help in 
clarifying the issue. This kind of estimate can be performed only in the main
body of the galaxy, where the stellar density is sufficient to allow the
derivation of well populated Color Magnitude Diagrams (CMD) from the observation of
reasonably small fields ($\simgt 0.1$ deg$^2$).
In this Letter we use a new large photometric dataset of 
Sgr stars to obtain a robust estimate of the mean age of the population that
dominates the stellar content of Sgr, shedding a new light on the M giant
{\em conundrum}.
 
%
   \begin{figure}
   \centering
   \vskip 4truecm
      \caption{ --- SEE 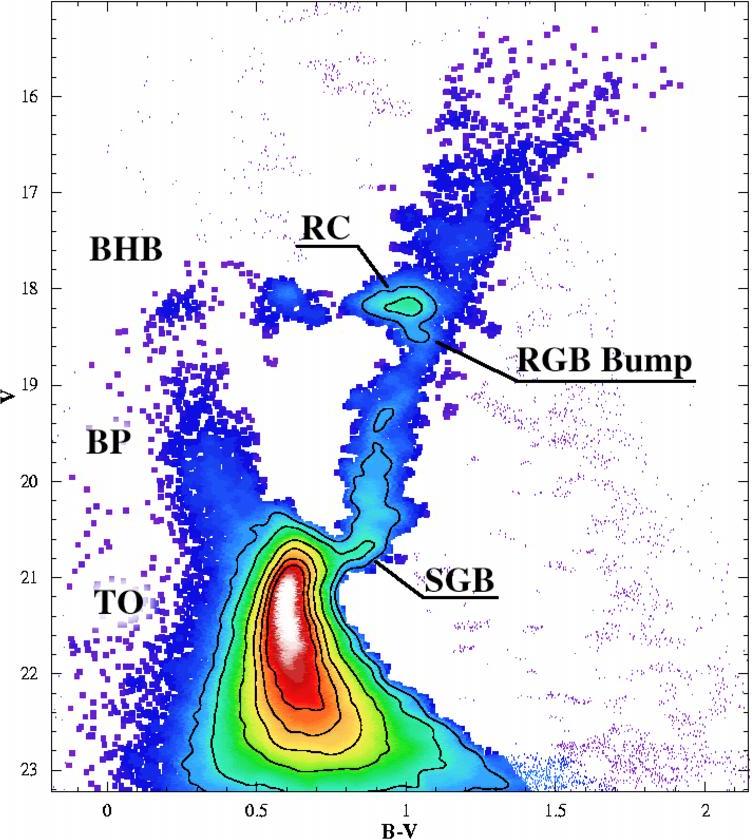 - POSTSCRIPT VERSION NOT INCLUDED ----
      Statistically decontaminated CMD of the Sgr34 Field.
      The color of the stars is coded according to the local density of stars on
      the diagram. A few representative density contours are also overplotted.
      Some remarkable features of the CMD are labeled: the Red
      Clump (RC), the Blue Horizontal Branch (BHB), the RGB Bump, the Blue Plume
      (BP), the Main Sequence Turn Off point (TO) and the Sub Giant Branch
      (SGB). The spurious residuals of the statistic decontamination process
      (mainly due to edge effects) to the lower left and the upper right of the
      main locus of the Sgr populations in the CMD have been plotted as 
      small points to provide a clearer view of the most significant parts of
      the diagram.
              }
         \label{fig1}
   \end{figure}
%


\subsection{The dataset}

We have obtained B,V,I photometry of a  $1\degr
\times 1\degr$ field located at (l,b) $\simeq(6.5\degr,-16.5\degr$), $\sim
2\degr$ to the East of the galaxy center, along the major axis (Sgr34
Field, see Bellazzini et al. \cite{BFBa}, hereafter BFBa). The field was imaged
with a mosaic of four pointings of the WFI camera mounted at the ESO/MPI 2.2m
telescope at La Silla, Chile. The data reduction was performed as in Monaco et
al. (\cite{lbump}, hereafter Mo02) and the  absolute photometric calibration 
was achieved with repeated observations of Landolt's (\cite{land}) 
standard fields. All the
details of the data acquisition and reduction will be described in a future
contribution (Bellazzini et al., in preparation).  
A $0.5\degr\times 0.5\degr$ field sampling the Galactic
population at similar angular distance from the Galactic Center (Gal Field, at 
(l,b)$\simeq(-6.0\degr,-14.5\degr$)) was also observed with the same camera, to
perform the statistical decontamination of the Sgr34 CMD
from the foreground/background Galactic Stars  (see
BFBb). The interstellar reddening was interpolated for each star from the
Schlegel et al. (\cite{cobe}) maps and corrected according to Bonifacio et al.
(\cite{bonir}). We found that the reddening variation over the considered
fields are negligible (with standard deviations $<0.01$ mag) and we adopted the
average reddening values $E(B-V)=0.116$ for Sgr34 and $E(B-V)=0.096$ for Gal
Field. 

The statistically decontaminated CMD obtained from the above described datasets
is presented in Fig.~1. The statistical decontamination has been performed as 
in BFBb. The large samples available for both Sgr34 (more than 300000 stars)
and for Gal Field ($\simeq 57000$ stars) ensure a reliable and clean recovery
of all the main features of the CMD, most of which have been labeled in Fig.~1
according to the nomenclature introduced in BFBa and Mo02.  
In the present context, the most  relevant features of Fig.~1
are the well identified TO point at $V_{TO}=21.4\pm 0.15$ and the 
single, narrow and well defined SGB that is clearly visible at
$V\simeq 20.8$ and $B-V\simeq 0.85$. In the following we will use these
observables to constrain the mean age of Sgr. 

   \begin{figure*}
   \centering
   \includegraphics[width=11.5cm]{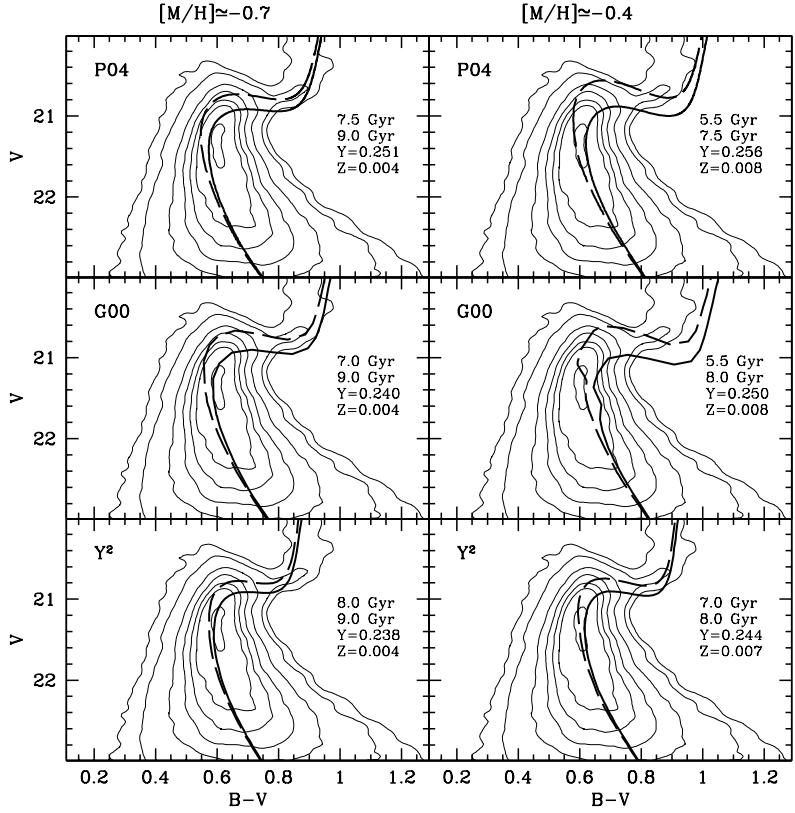}
   \caption{Isochrone fitting of the TO/SGB region of the Sgr34 CMD, here 
   represented with the isodensity contours shown in Fig.~1. 
   The isochrones plotted as continuous lines provides the best fit to the
   TO luminosity, those plotted as dashed lines provides the best fit to the 
   luminosity of the SGB.
   The left panels show the comparison with Z=0.004 isochrones,
   the right panel show the comparison with Z=0.007-0.008 isochrones. Isochrones
   from the P04 set (upper panels), G00 set (middle panels), 
   and Y$^2$ set (lower panels) are considered. 
   The age, metallicity and helium content (Y) of the adopted 
   isochrones are reported in each panel.}
              \label{isoc}%
    \end{figure*}
%

\section{The mean age of Sgr dSph}

The availability of a large and complete sample of Horizontal Branch (HB) stars
allows an accurate census of the stars older than $\sim 1-2$ Gyr in a given
stellar system. With this method Monaco et al. (\cite{lbhb}) established that
$\sim$ 12\% of Sgr stars in this age range belong to a population with age
$\simgt 10$ Gyr and metallicity [Fe/H]$\simlt -1.3$ (traced by BHB and RR Ly
stars, a result confirmed also with the present sample). Apart from these old
and metal poor stars and a very sparse (possibly young) population traced by
the BP accounting for $\simlt$ 6\% of the whole stellar content (BFBa), more 
than 80\% of Sgr stars belong to a metal-rich and intermediate-old age 
population (Pop~A, after BFBb and Mo02; see also LS00 and references therein).
 Observations of the RGB in the Sgr core (Cole \cite{cole}, M03) show 
that the 
dominant RGB sequence -- which must correspond to Pop A -- includes a 
prominent population of M giants at its tip.  It is logical to conclude 
that the M giants observed in the Sgr tails come from this population
(rather than from, e.g., the much smaller and presumably even younger 
BP population). Thus, constraining the age of Pop A would then mean to 
constrain the age of M giants in the main body, and therefore the Stream, 
of Sgr.
The compact morphology of the Red Clump and of the TO region and the presence
of a single and well defined RGB Bump (Mo02) strongly suggest that Pop~A stars 
spans a relatively limited age and metallicity range. The most likely SFH
proposed by BFBb indicates that the episode that lead to the formation of Pop~A
lasted $\sim 3$ Gyr. 
Joining the samples of abundances of {\em bona fide} 
Sgr stars obtained with high resolution spectroscopy by 
Bonifacio et al. (\cite{boni4}, B04) and by Monaco et al. (\cite{luves}, M05)
 we obtain a metallicity distribution from 27 stars
that shows a strong peak around $[M/H]\simeq -0.5$\footnote{All over this
letter we will use the {\em global metallicity}  parameter as defined by
Salaris, Chieffi \& Straniero (\cite{scs93}), 
$[M/H]=[Fe/H]+log(0.638\times10^{[\alpha/Fe]}+0.362)$. The average of the
[Mg/Fe] and [Ca/Fe] abundances are used as a proxy of $ [\alpha/Fe]$ (see
Monaco et al. \cite{luves}).To obtain a final sample as homogeneous as possible,
we have recomputed the abundances of B04 stars from the published 
equivalent widths, adopting the same temperature scale (i.e. photometry and 
reddening) and atomic parameters used by M05. 
This lead to only minor revisions of the B04 estimates 
(i.e. an average shift  of $-0.17 \pm 0.04$ dex), 
removing the small offset between the two scales noted in M05.}. 
The average and standard deviation of the
distribution for $[M/H]\ge -1.2$ (26 of the 27 stars of the sample) are
respectively $[M/H] = -0.55$ and $\sigma=0.22$ dex. A preliminary analysis of
the much larger (114 stars) sample described by 
Zaggia et al. (\cite{za}) fully confirms the essence of this result: the
distribution shows a clear peak at $[M/H] \sim -0.6$ with the same $\sim
0.25$ dex dispersion. 
All the above considerations imply that a quantity as the ``mean age'' 
of Pop~A is well defined and can be estimated from the CMD of Fig.~1. 

In Fig.~2 we compare the TO-SGB region of the decontaminated CMD with
theoretical isochrones.
To account for the
metallicity range spanned by the bulk of Pop~A stars we consider isochrones
with Z=0.004 and Z$\simeq$ 0.008. To limit the sensitivity on the various
assumptions influencing the stellar evolution models we perform the comparison
with three different sets of isochrones, namely those by Pietrinferni et al.
(\cite{P04}, P04), by Girardi et al. (\cite{G00}, G00), and the Yale-Yonsei set
(Yi et al. \cite{Y01}, $Y^2$). The distance modulus $\mu_0=(m-M)_0$=17.10, 
recently obtained by Monaco et al. (\cite{ltip}) is adopted. 

The luminosity of the TO point is the most reliable age indicator from a
theoretical point of view. However, Pop~A stars of different
ages and metallicities, having different TO luminosity and colors, contribute
to the definition of the TO point observed in Fig.~1. On the other hand
the observed CMD shows that the vast majority Pop~A stars have essentially the 
same luminosity during the SGB phase (see also Fig.~8 of BFBb, and the 
discussion reported there). 
Hence, in spite of the larger sensitivity
of the SGB morphology to chemical composition and to subtleties of the stellar
models, the single and narrow SGB may also provide a very useful constraint on 
the mean age of Pop~A, since isochrones of any age and metallicity must 
reproduce this feature to provide an acceptable fit of the observed CMD.  
In each panel of Fig.~2 we plot both the isochrone that optimize the fit of the
TO luminosity (continuous lines) and the isochrone that optimize the fit of the
SGB luminosity (dashed lines). While the fit of the SGB provides systematically
younger age estimates, the difference between the results of the two 
fits is always $\le 2.0$ Gyr, and typically $\simeq 1.0$ Gyr. Hence the two
methods gives essentially the same answers, within the uncertainties of each
single estimate. 

The main conclusion that can be drawn from Fig.~2 is that {\em the mean age 
of Pop~A is $\sim 8.0 \pm 1.5$ Gyr for Z=0.004 and $\sim 6.5 \pm 1.5$ Gyr 
for Z=0.008, independently of the considered set of isochrones}. We note that
the Z=0.004 isochrones provide an overall better fit to the observed CMD, 
supporting the ``older'' age estimate.
To provide a link to a more familiar age scale, we derived the age of the
globular cluster 47~Tuc (which has $[M/H]=-0.66$, Ferraro et al. \cite{f99})
comparing the Z=0.004 isochrones with the CMD by Bellazzini et al.
(\cite{mtip}).  All the considered isochrone sets provide an age of 12-13 Gyr
for 47~Tuc, confirming that the Pop~A is $\sim 5-7$ Gyr younger, in average, 
than the cluster (BFBb, Mo02).

   \begin{figure}
   \centering
   \includegraphics[width=7.cm]{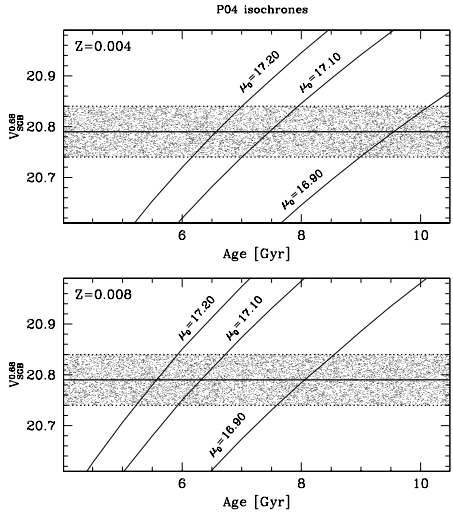}
      \caption{Effect of varying the assumptions on the distance modulus of Sgr
      on the mean age estimates derived from the magnitude of the SGB at
      $(B-V)_0=0.68\pm 0.02$ ($V_{SGB}^{0.68}$). The continuous horizontal line
      marks the observed $V_{SGB}^{0.68}$ level, the dotted lines enclose the
      range of $V_{SGB}^{0.68}$ allowed by the observations, given the 
      uncertainties (hatched).
      The curves labeled with different distance moduli ($\mu_0$) are the
      theoretical predictions for $V_{SGB}^{0.68}$ as a function of age computed
      from the P04 isochrones.
      Both the Z=0.004 (upper panel) and Z=0.008 (lower panel) cases are
      considered.
              }
         \label{fig3}
   \end{figure}
%

To consider the impact of different assumptions about the distance modulus on
our age estimate we proceeded as follows: we computed the average V magnitude of
the stars in the most horizontal region of the SGB, i.e. $0.66\le (B-V)_0\le
0.70$, finding $V_{SGB}^{0.68}=20.79\pm 0.05$; we derived the same parameter
from the Z=0.004 and Z=0.008 isochrones of the P04 set, so obtaining a
calibration of $V_{SGB}^{0.68}$ as a function of age. The theoretical
$V_{SGB}^{0.68}$ vs. Age relations 
were corrected for the reddening of Sgr34 and for different
assumptions on the distance modulus, spanning from the shortest
($\mu_0=16.90$, Alard \cite{ala}) to the longest ($\mu_0=17.19$, LS00) 
scales found in the literature (see Monaco et al. (\cite{ltip}) for a 
discussion).
In Fig.~3 we compare the observed value of $V_{SGB}^{0.68}$ with the theoretical
predictions for different distance moduli. It is evident that even with the most
extreme assumptions ($\mu_0=17.20$ and Z=0.008) the lowest mean age allowed by the
data is $>5.0$ Gyr while the range of the best-fit estimates is 
5.5 Gyr$\le$ age $\le$ 9.5 Gyr. Note that a similar experiment performed
using the magnitude of the TO as age indicator would lead to slightly older
average ages.

 The different assumptions on distance
modulus are at the origin of the younger ``best-fit'' ages found by LS00 with
respect to the present analysis. Moreover, it should be considered that the AMR
adopted by M03 for their estimate of the age of M giants in the Sgr Stream was
obtained (by LS00) when essentially no spectroscopic estimate of the metallicity
of Sgr stars was available. Hence, while it may provide useful general
constraints on the chemical evolution history of Sgr (see LS00), the inherent 
uncertainties are probably too large to use it to convert metallicities into 
ages. For
example, if we convert spectroscopic [Fe/H] (from the M05+B04 sample discussed
above) into ages using the LS00 AMR we find that 70\% of the stars have an age
$\le 3.0$ Gyr. This is in sharp contrast with the CMD of Sgr shown in Fig.~1
and 2, above: isochrones of ages $\le 3.0$ Gyr would provide an exceedingly 
bad fit to the observed TO and SGB. 
Finally, the conversion of metallicities into ages with a
non-linear AMR may propagate acceptable uncertainties in metallicity into large
- and metallicity dependent - errors in age. With the AMR of LS00, an error 
of $\pm 0.2$ dex in metallicity lead to a maximum error in the derived age of
$\sim \pm 3.0$ Gyr, at $[Fe/H]\sim -0.6$.

It is important to stress, here, that we are {\em not} claiming that {\em all} 
Pop~A stars are older than 5 Gyr. It is very likely that younger stars are
also there and that Pop~A stars follow an AMR (Montegriffo et al. \cite{paolo},
LS00). The key result of this
Letter is that the majority of Pop~A stars have an age sufficiently large to
allow them (and the associated M giants) to populate the observed Stream, 
in agreement with the current best-fit models of the disruption of the Sgr 
galaxy.

\section{Conclusions}

The above described results robustly establishes that an age of 5 Gyr is not 
an upper limit for the bulk of Pop~A (and, consequently, of M giants in the 
main body of Sgr and in the Stream) as assumed by M03, but, in fact, a 
strong {\em lower} limit to the age of most of these stars. 
Moreover, independently of the adopted theoretical models and of the
assumed distance, the best fit age for this population lies in the range
$5.5\le$ age $\le 9.5$ Gyr, and our preferred solution (Z=0.004 and
$\mu_0=17.10$, averaging over the results from the different isochrones 
sets) is $\langle$age$\rangle_{PopA} = 8.0\pm 1.5$ Gyr.
This {\em evolutionary timescale} is now comfortably larger than the {\em
dynamical timescale} provided by realistic simulations of the formation of the
Sgr Stream (Law et al. \cite{law}), hence the {\em M giant conundrum} appears
to be solved.

\begin{acknowledgements}
This research is partially supported by the INAF-PRIN2005 grant assigned to the
project "A hierarchical merging tale told by stars: motions, ages and chemical
compositions within structures and substructures of the Milky Way".
The financial support of MIUR is also acknowledged. We are grateful
to the Referee (S.R. Majewski) for insightful discussions and for very
useful suggestions that improved the overall quality of the paper.
 \end{acknowledgements}

\end{document}